\documentclass[11pt,a4wide]{article}

\usepackage{amsmath}
\usepackage{amsfonts}
\usepackage{amsbsy}
\usepackage{hyperref}
\usepackage{epsfig}
\usepackage{latexsym}
\usepackage{color,ulem}

\advance \topmargin by -\headheight
\advance \topmargin by -\headsep
\evensidemargin \oddsidemargin
\advance\hoffset by -3mm  
\advance\voffset by  -3mm  
\def\bbbc{{\mathchoice {\setbox0=\hbox{$\displaystyle\rm C$}\hbox{\hbox
to0pt{\kern0.4\wd0\vrule height0.9\ht0\hss}\box0}}
{\setbox0=\hbox{$\textstyle\rm C$}\hbox{\hbox
to0pt{\kern0.4\wd0\vrule height0.9\ht0\hss}\box0}}
{\setbox0=\hbox{$\scriptstyle\rm C$}\hbox{\hbox
to0pt{\kern0.4\wd0\vrule height0.9\ht0\hss}\box0}}
{\setbox0=\hbox{$\scriptscriptstyle\rm C$}\hbox{\hbox
to0pt{\kern0.4\wd0\vrule height0.9\ht0\hss}\box0}}}}

\makeatletter
\@addtoreset{equation}{section}
\makeatother

\begin{document}

\hfuzz=100pt \title{{\Large \bf{{Kottler Spacetime in Isotropic Static Coordinates}}}}

\author{
R Solanki \footnote{E-mail: rahulkumar.solanki@utdallas.edu}
\\
\\{Department of Physics}
\\ {University of Texas at Dallas}
\\ {Richardson, TX 75080, USA}}
\date{\today}
\maketitle

\begin{abstract}
\noindent
The Kottler spacetime in isotropic coordinates is known where the metric is time-dependent. In this paper, the Kottler spacetime is given in isotropic static coordinates (i.e., the metric components are time-independent). The metric is found in terms of the Jacobian elliptic functions through coordinate transformations from the Schwarzschild-(anti-)de Sitter metric.
In canonical coordinates, it is known that the unparameterized spatially projected null geodesics of the Kottler and Schwarzschild spacetimes coincide.
We show that in isotropic static coordinates, the refractive indices of Kottler and Schwarzschild are not proportional, yielding spatially projected null geodesics that are different.
\end{abstract}

\section{Introduction}
The McVittie metric \cite{McVittie:1933zz} is the Kottler spacetime in isotropic coordinates in which the metric components are time-dependent:
\begin{equation}
  \label{McVittie}
  ds^2 = -\left(\frac{1-\mu}{1+\mu}\right)^2 d\bar{t}^{\;2} + (1+\mu)^4\; a(\bar{t})^2 \left[d\bar{r}^{\;2} +\bar{r}^2 (d\theta^2+\sin^2\theta d\phi^2)\right],
\end{equation}
where
\begin{equation}
  \label{McVittie_more}
  a(\bar{t})=e^{H\bar{t}}=e^{\sqrt{\frac{\Lambda}{3}}\;\bar{t}} \quad \text{and} \quad \mu=\frac{M}{2a\bar{r}}.
\end{equation}
For $\Lambda=0$ (i.e., $a(t)=1$), this line element reduces to the Schwarzschild spacetime in isotropic coordinates and
for $M=0$, to the Friedmann-Lema\^{i}tre-Robertson-Walker (FLRW), i.e., $\Lambda$ only universe.
Moreover, for $\mu << 1$, the McVittie metric turns to perturbed the FLRW in a Newtonian gauge \cite{Dodelson:2003ft}.
Under the transformation
\begin{equation}
  \label{McVittie_to_Kottler}
  r=a\bar{r} (1+\mu)^2 \;,\; t=\bar{t}+ \xi(r) \quad \text{where} \quad \frac{d\xi}{dr}=\frac{rH}{f(r)\sqrt{1-{2M/r}}},
\end{equation}
the McVittie metric transforms into the Kottler metric in canonical coordinates (see equation (\ref{Kottler_canonical}) below).

In this paper, the Kottler spacetime is given in isotropic coordinates where the metric components are \textit{time-independent}, i.e., in isotropic static coordinates (because $g_{0i}=0$).
The two cases for which the spacetime is static are the region between two horizons when $0<\Lambda<1/9M^2$ and the region beyond the single horizon when $\Lambda<0$.
The case with two horizons is presented first, followed by one horizon.
The metric tensor in isotropic coordinates $\{t,\rho,\theta,\phi\}$ is obtained by transforming from canonical coordinates $\{t,r,\theta,\phi\}$, and as it turns out, the transformation $r=r(\rho)$ follows Jacobian elliptic functions.
In isotropic coordinates, the spatial metric is conformal to the Euclidean metric, whereas in canonical coordinates, the circumference of a circle centered at the origin is $2\pi r$.
Throughout the paper, the dot denotes differentiation with respect to the argument.
In order to distinguish between the limit and the variable of integration, prime is used for the latter.
Moreover, Greek indices represent spacetime coordinates (where $x^0\equiv t$) and Roman indices represent spatial coordinates.

The metric in isotropic static coordinates has several advantages.
In static coordinates, the propagation of light follows the variational principle of Fermat, and in isotropic coordinates, the spatial metric shares the same conformal properties as the Euclidean.
Moreover, put together, the light propagation can be mimicked by an optical medium in ordinary geometric optics with an appropriate index of refraction \cite{Perlick:2010zh}.
For the line element in static coordinates
\begin{equation}
  \label{line_element_static}
  ds^2\equiv g_{\mu\nu}dx^{\mu}dx^{\nu}=g_{00}dt^2+g_{ij}dx^{i}dx^{j}=-g_{00}\left(\tilde{g}_{\mu\nu}dx^{\mu}dx^{\nu}\right),
\end{equation}
the spatial trajectories of null geodesics (i.e., projection of null geodesics on a constant time hypersurface) are given by the geodesics of the optical metric\footnote{In this paper, we consider light rays propagating along the null geodesics of the spacetime metric. Furthermore, the optical metric $n_{ik}$ is of the Riemannian signature.

If the propagation of light is influenced by an isotropic non-dispersive optical medium, then the light rays follow the null geodesics of another Lorentzian metric (see e.g., \cite{Perlick_book} and references therein). In literature, this metric of the Lorentzian signature is also called the optical metric. Therefore, to avoid potential confusion, $n_{ik}$ will be referred to as the Fermat metric henceforth.} (or Fermat metric) \cite{Abramowicz88, Gibbons:2008ru, Perlick:2010zh}
\begin{equation}
  \label{optical_metic_static_spacetime}
  ds^2_{opt}=-\frac{g_{ij}}{g_{00}}dx^{i}dx^{j}=n_{ij}dx^{i}dx^{j}.
\end{equation}
One can prove this using the following property of conformal transformations (for a proof using Fermat's principle, see for example, page 273 of Ref.\cite{Landau:1982dva}).
Null geodesics remain invariant under conformal transformations; however, their parameterization changes from affine to non-affine (see Appendix D of Ref.\cite{Wald:1984rg}).
For the metric $\tilde{g}_{\mu\nu}$, since $\tilde{g}_{00}=-1$ and $\tilde{g}_{0i}=0$, the Christoffel symbols $\tilde{\Gamma}^{\;\mu}_{\;0\nu}$ vanish.
Therefore, the spatial part of null geodesics\footnote{The temporal part, since $\tilde{\Gamma}^{\;0}_{\;ik}=0$, is given by, $dt/d\tilde{\lambda}=$ constant. Thus, $\tilde{\lambda}$ is affine parameter to both, null geodesics of $\tilde{g}_{\mu\nu}$ and geodesics of $n_{ik}$, since $$\tilde{g}_{\mu\nu}\frac{dx^\mu}{d\tilde{\lambda}}\frac{dx^\nu}{d\tilde{\lambda}}=0 \implies \frac{dt}{d\tilde{\lambda}}=n_{ik}\frac{dx^i}{d\tilde{\lambda}}\frac{dx^k}{d\tilde{\lambda}}=\text{constant.}$$} follow
\begin{equation}
  \label{spatial_part_null_conformal}
  \frac{d^2x^i}{d\tilde{\lambda}^2}+\tilde{\Gamma}^{\;i}_{\;jk}\frac{dx^j}{d\tilde{\lambda}}\frac{dx^k}{d\tilde{\lambda}}=0,
\end{equation}
which is precisely the geodesic equation of the Fermat metric $n_{ij}$.
In other words, the spatial trajectories of null geodesics are the geodesics of the Fermat metric $g_{ij}/|g_{00}|$, not of spatial metric $g_{ij}$ \cite{Landau:1982dva}.

For the Schwarzschild and Kottler spacetimes in canonical coordinates, it is known that the spatial trajectory of null geodesics satisfies the identical second-order ordinary differential equation \cite{Islam:1983rxp}.
This differential equation, however, represents unparameterized curves ($r=r(\phi)$ in this case, if the equatorial plane is considered).
Therefore, the proper length and conformal properties such as angles are not identical.
For example, in three dimensions, the unparameterized geodesics of both Euclidean and Beltrami metrics\footnote{The line element for one half of the sphere is given by, $$dl^2=\frac{dr^2}{(1+r^2/R^2)^2}+\frac{r^2 \left(d\theta^2+\sin^2\theta\,d\phi^2\right)}{(1+r^2/R^2)}.$$} are straight lines. However, parameterizing these geodesics by arc length reveals the difference \cite{Gibbons:2008ru}.
This property that the unparameterized geodesics of the Fermat metrics coincide is known as projective equivalence.
If two Fermat metrics $n_{ij}$ and $\bar{n}_{ij}$ are projectively equivalent, then the projective curvature tensors $W^{i}_{\; klm}$ and $\bar{W}^{i}_{\; klm}$ are equal. Here, the Weyl projective curvature tensor, constructed from the Fermat metric $n_{ij}$ (of the Riemannian signature), is given by  \cite{Eisenhart1926}
\begin{equation}
  \label{projective_curvature_tensor}
  W^{i}_{\;klm}=R^{i}_{\;klm}+\frac{1}{2}(\delta^{i}_{m}R_{kl}-\delta^{i}_{l}R_{k m}).
\end{equation}
In general, for an ${m}$-dimensional Riemannian manifold, the factor of $1/2$ is replaced by $1/(m-1)$.\footnote{The Riemannian metrics $n_{ij}(x^k)$ and $\bar{n}_{ij}(x^k)$ are projectively equivalent if and only if their associated Christoffel symbols are related by
\begin{equation}
   \label{projective_Christoffel}
   \bar{\Gamma}^{i}_{kl}=\Gamma^{i}_{kl}+\delta^{i}_{k}a_{l}+\delta^{i}_{l}a_{k}\;,
\end{equation}
where $a_k$ is a covariant vector \cite{Eisenhart1926}.
For an arbitrary $a_k$, the changes in the Christoffel symbols above are called projective transformations.
In other words, projective transformations preserve unparameterized geodesics, and the projective curvature tensor $W^{i}_{\;klm}$ is invariant under these transformations \cite{SS}.
This tensor vanishes identically in two dimensions.
In dimensions greater than $2$, a metric $n_{ij}(x^k)$ is projectively flat (i.e., there exists a projective transformation through which one can obtain $\bar{n}_{ij}(x^k)$ such that $\bar{R}_{iklm}=0$) if and only if $W^{i}_{\;klm}$ vanishes \cite{SS}.
Basically, unparameterized geodesics are straight lines for a projectively flat metric.
Conformal transformations, $\tilde{n}_{ij}(x^l)=\exp\left({2\omega}\right) {n}_{ij}(x^l)$ where $\omega=\omega(x^l)$, on the other hand, preserve angles in corresponding directions at corresponding points \cite{Eisenhart1926}.
Here, the associated Christoffel symbols are related by
\begin{equation}
   \label{conformal_Christoffel}
   \tilde{\Gamma}^{i}_{kl}=\Gamma^{i}_{kl}+\delta^{i}_{k}\partial_l \omega+\delta^{i}_{l}\partial_k \omega-g_{kl}\partial^i \omega \;,
\end{equation}
where $\partial_k\equiv \partial/\partial x^k$. 
The conformal tensor $C^{i}_{\;klm}$ is invariant under conformal transformations and vanishes identically in three dimensions \cite{Eisenhart1926}.
For the Lorentzian metrics, the unparameterized null geodesics remain invariant under conformal transformations; although relationship (\ref{projective_Christoffel}) is not satisfied, the last term in (\ref{conformal_Christoffel}) vanishes due to the null condition when (\ref{conformal_Christoffel}) is substituted in the null geodesic equation.}
In this paper, it is shown that for the Schwarzschild and Kottler spacetimes in isotropic coordinates, the conformal properties of the corresponding Fermat (and spatial) metrics are the same, but the projective equivalence no longer exist (i.e., opposite example to Ref.\cite{Gibbons:2008ru}).\footnote{Consider a metric $g_{\mu\nu}$ in static coordinates (i.e. $\partial_0g_{\mu\nu}=0$ and $g_{0i}=0$) with the Fermat metric $n_{ik}$. Under the coordinate transformations $x'^{\mu}=x'^{\mu}(x^{\nu})$, $$g'_{00}=g_{00},\quad n'_{ik}=\frac{\partial x^{l}}{\partial x'^{i}}\frac{\partial x^{m}}{\partial x'^{k}} n_{lm},\quad W'^{i}_{\,\,klm}=\frac{\partial x'^i}{\partial x^a}\frac{\partial x^b}{\partial x'^k}\frac{\partial x^c}{\partial x'^l}\frac{\partial x^d}{\partial x'^m} W^{a}_{\,\,bcd},$$ if $g'_{\mu\nu}$ is also static. If two Fermat metrics $n_{lm}$ and $\bar{n}_{lm}$ are projectively equivalent, then $$W^{a}_{\,\,bcd}=\bar{W}^{a}_{\,\,bcd}.$$ However, this does not necessarily mean that $W'^{i}_{\;\;klm}$ and $\bar{W'}^{i}_{\,\,klm}$ are equal.
(If they vanish, on the other hand, the equality holds in all static coordinates because if a tensor vanishes in one coordinate system, then it vanishes in all because of the tensor transformation rule. The projective curvature tensor vanishes for a constant curvature space \cite{Gibbons:2008ru}.) This can be explained, for example, through transformations between the isotropic $\{t, \rho, \theta, \phi\}$ and canonical $\{t, \rho, \theta, \phi\}$ coordinates. In general, the transformation between the radial coordinates, $r=r(\rho)$ or $\rho=\rho(r)$, is different for different spacetimes.}
In other words, in isotropic static coordinates, the unparameterized spatially projected null geodesics of the Kottler spacetime are $\Lambda-$dependent, and the angles measured by stationary observers are the same as coordinate angles (locally). Thus, the isotropic static form of the Kottler spacetime is more suitable for studying gravitational lensing using quasi-Newtonian approximations (e.g., the lens equation relating source and image positions; see \cite{EFS} and references therein).

The paper is organized as follows.  
In Section $2$, the equivalence between the spatially projected null geodesics of the Kottler and Schwarzschild is extended from unparameterized to affinely parameterized curves.
The analogy between the index of refraction (in standard geometric optics) and the projective equivalence (in isotropic coordinates) is explored in Section $3$.
The problem of finding a static isotropic metric is systematically developed in Section $4$.
In Section $5$, for $0<\Lambda<1/9M^2$, the metric is calculated in terms of Jacobian elliptic functions.
Similarly, for $\Lambda<0$, the solution is presented in section $6$.
In both these sections, the index of refraction is plotted and the Schwarzschild limit is discussed.

\section{Remark on Parameterization}
The unparameterized geodesics of projectively equivalent Fermat metics coincide, but affine (e.g., arc-length) parameterization differs \cite{Eisenhart1926, Gibbons:2008ru}.
However, the spatial trajectory of affinely parameterized null geodesics (of the corresponding spacetime metrics) can be identical; an example is the Kottler and Schwarzschild spacetimes in canonical coordinates. Here, the null geodesics of the Kottler metric (\ref{Kottler_canonical}) in the equatorial plane can be derived, for example, using the same approach as for the Schwarzschild metric (see Section $6.3$ of Ref.\cite{Wald:1984rg}): 
\begin{gather}
  \left(1-\frac{2M}{r}-\frac{\Lambda r^2}{3}\right)\dot{t}=E, \\
  r^2\dot{\phi}=J, \\
    \dot{r}^2=\left(E^2+\frac{\Lambda}{3}J^2\right)-\frac{J^2}{r^2}\left(1-\frac{2M}{r}\right) \label{null_constant},
\end{gather}
where $E$ and $J$ are the constants of motion, energy and angular momentum of the photons respectively. 
Moreover, $\dot{t}=dt/d\lambda$ where $\lambda$ is affine parameter and equation (\ref{null_constant}) is the null condition in terms of the constants of motion.
If the affine parameter is rescaled such that $\dot{x}^{\mu}$ is the wave four-vector, then $E$ becomes the frequency of light rays \cite{Landau:1982dva}. 
As can be seen, $r(\lambda)$ and $\phi(\lambda)$ if the frequency is modified to
\begin{equation}
    \bar{E}=\sqrt{E^2+\frac{\Lambda}{3} J^2},
\end{equation} 
are the same as those in Schwarzschild.
In other words, the spatial trajectory of affinely parameterized null geodesics of Kottler and Schwarzschild spacetimes (in canonical coordinates) coincide but the frequency of light rays differs.

\section{Index of Refraction and Projective Equivalence}
In ordinary optics, the index of refraction represents the ratio between the speed of light in vacuum and in the optical medium.
Therefore, for an isotropic optical medium with an index of refraction $n(\rho)$ the time it takes for a light ray to traverse spatial distance (say between spatial points 1 and 2) is given by,
\begin{equation}
  \label{light_travel_time}
  t=\int^{2}_{1} n(\rho)dl
\end{equation}
where $dl^2=d\rho^2+\rho^2 \left(d\theta^2+\sin^2{\theta}\,d\phi^2\right)$ and $n(\rho)dl$ is the optical length.
Moreover, according to Fermat's variational principle, the light ray follows spatial trajectory (between points 1 and 2) for which the optical length is stationary, i.e., $\delta t=0$ \cite{Lakshmi2002}.
This is equivalent to finding the geodesics of a curved space (Riemannian) with the line element,
\begin{equation}
  \label{optical_metric}
  dl^2_{opt}=n^2(\rho) \left(d\rho^2+\rho^2 d\Omega^2 \right),
\end{equation}
where $d\Omega^2=d\theta^2+\sin^2{\theta}\,d\phi^2$.

The spherically symmetric static spacetime in isotropic coordinates can be written as
\begin{equation}
\label{isotropic_metric}
ds^2= F(\rho) \left[-dt^2+n^2(\rho)(d\rho^2+\rho^2 d\Omega^2) \right].
\end{equation}
The associated Fermat metric, from equation (\ref{optical_metic_static_spacetime}), is given by equation (\ref{optical_metric}).
Therefore, the metric function $n(\rho)$ is analogous to the index of refraction of a spherically symmetric optical medium in ordinary optics (see \cite{Perlick:2010zh} and references within).
The metric of equation (\ref{isotropic_metric}) follows the same relationship (\ref{light_travel_time}), since the spacetime interval vanishes for the null path.
And for the metric conformal to it (conformal factor $F(\rho)$), $t$ is the proper time and $n(\rho)dl$ is the proper length.
The projective equivalence between two Fermat metrics with refractive indices $n(\rho)$ and $\bar{n}(\rho)$ can be determined directly from the unparameterized geodesic equation (see Appendix \ref{Light Rays in Isotropic Coordinates}):
\begin{equation}
  \label{null_trajectory}
  \frac{d^2\rho}{d\phi^2}=\left(\frac{d\rho}{d\phi} \right)^2 \left[\frac{\dot{n}}{n}+\frac{2}{\rho} \right]+\rho+\rho^2 \left(\frac{\dot{n}}{n}\right).
\end{equation}
Here, the coefficients of the differential equation are invariant, if and only if the ratio $\dot{n}/n$ remains the same, which after integration suggests that $\bar{n}(\rho)/n(\rho)=$constant.
Thus, in isotropic (static) coordinates, two Fermat metrics are projectively equivalent if and only if their refractive indices are proportional to each other, with the same proportionality constant everywhere \cite{Akbar_Behshid2021}.

\section{From Canonical to Isotropic coordinates}
The Kottler spacetime in Schwarzschild canonical coordinates is given by
\begin{equation}
  \label{Kottler_canonical}
  ds^2=-f(r)dt^2+\frac{dr^2}{f(r)}+r^2d\Omega^2,
\end{equation}
where $f(r)=\left(1-\frac{2M}{r}-\frac{\Lambda r^2}{3}\right)$.
For $\Lambda>0$, it is also called the Schwarzschild-de Sitter metric and for $\Lambda<0$, Schwarzschild-anti-de Sitter metric.
Transformation to isotropic coordinates (i.e., equation (\ref{isotropic_metric})) yields
\begin{equation}
  \label{transformation_rule}
  \frac{dr}{d\rho}=\frac{r}{\rho}\sqrt{f(r)},\quad n(\rho)=\frac{r}{\rho}\frac{1}{\sqrt{f(r)}},\quad F(\rho)=f(r).
\end{equation}
The variables in the first equation (ordinary differential) are already separated with the term $r\sqrt{f(r)}$ polynomial of degree four under the square root.
If the polynomial has no multiple roots, then the solution of this equation reduces to an elliptic integral.
The transformation between radial coordinates $\rho=\rho(r)$ is given by Legendre's canonical elliptic integral of the first kind (see Appendix \ref{elliptic}).
However, since we are interested in $r=r(\rho)$, the solution reduces to the elliptic functions of Jacobi, obtained by the inversion of elliptical integrals of the first kind. Evaluation of the integral requires knowing the roots of the polynomial (including the static region and horizons to specify the range).
Of course, $r=0$ is one root of the polynomial,
\begin{equation}
  r^2f(r)=-\frac{\Lambda}{3}r\left(r^3-\frac{3r}{\Lambda}+\frac{6M}{\Lambda}\right)=-\frac{\Lambda}{3}r\bar{f}(r).
\end{equation}
If $r_1$, $r_2$ and $r_3$ are the roots of cubic equation $\bar{f}(r)=0$, then
\begin{equation}
  \label{root_properties}
  r_1 r_2 r_3=-\frac{6M}{\Lambda},\quad r_1+r_2+r_3=0,\quad r_1r_2+r_2r_3+r_3r_1=-\frac{3}{\Lambda}.
\end{equation}
Here, the corresponding values in isotropic coordinates are taken as $\rho_1$, $\rho_2$ and $\rho_3$.
Additionally, if (see e.g., \cite{Stegun} for the solution of the cubic equations or Appendix \ref{cubic})
\begin{equation}
  \label{condition_roots}
  \left(\frac{9M^2}{\Lambda^2}-\frac{1}{\Lambda^3}\right) \quad
   \left\{
   \begin{aligned}
     &>0, \text{ i.e., } \Lambda<0\;\text{or}\;\Lambda>\frac{1}{9M^2}, \text{ then a real root and a complex conjugate pair.} \\
     &=0, \text{ i.e., } \Lambda=\frac{1}{9M^2}, \text{ then all roots are real and at least two are equal.} \\
     &<0, \text{ i.e., } 0<\Lambda<\frac{1}{9M^2}, \text{ then all roots are real.}
   \end{aligned}
   \right.
\end{equation}
Let's consider each case separately (see e.g., section $5.2$ of Ref.\cite{Perlick:2010zh}).
\subsubsection*{$\Lambda<0$}
In this case, from equation (\ref{root_properties}), the product of all roots is positive. And since the product of the complex conjugate pair (e.g. $r_2$ and $r_3$) is positive, the real root ($r_1$) is positive.
Thus, there is one horizon at $r=r_1$ with a static region outside (for $r_1<r<\infty$, $f(r)>0$).\footnote{Since $\bar{f}(r_1)=0$, $\left(r_1-2M\right)=\Lambda r_1^3/3$. Thus, $0<r_1<2M$ since $\Lambda <0$.}

\subsubsection*{$0<\Lambda<1/9M^2$}
All roots are real in this case. The product of all roots is negative and the sum is zero. Therefore, one root (e.g. $r_1$) is negative and the rest are positive (take $r_2<r_3$).
Therefore, there are two horizons at $r=r_2$ and $r=r_3$ with a static region in between ($f(r)>0$ for $r_2<r<r_3$).\footnote{Here, $r_3>r_2>2M>0$, since $\Lambda>0$ and $\bar{f}(r_2)=\bar{f}(r_3)=0$. Furthermore, from (\ref{root_properties}), $$2\,r_3^3>r_2\,r_3(r_2+r_3)=\frac{6M}{\Lambda}>2\,r_2^3.$$Thus, $\left(\Lambda r_3^3/3\right)-M=r_3-3M>0$ and $\left(\Lambda r_2^3/3\right)-M=r_2-3M<0$. In other words, $r_1<0<2M<r_2<3M<r_3$.}

\subsubsection*{$\Lambda=1/9M^2$}
Since the sum of all roots vanishes, only two roots (e.g. $r_2$ and $r_3$) are equal.
And from the product of the roots, the remaining root ($r_1$) is negative, since $\Lambda>0$.
Thus, there is one horizon at $r=r_2=r_3$, however, there is no static region.

\subsubsection*{$\Lambda>1/9M^2$}
In this case, the real root (e.g. $r_1$) is negative, since the product of the roots is negative and the product of the complex conjugate pair ($r_2$ and $r_3$) is positive.
Since $f(r)\neq 0$ for $r>0$, there are no horizons.
Additionally, $f(r)<0$ for $r>0$, therefore, no static region exists.

Thus, only for the first two cases, static region exists.
The line element for second case (with two horizons) is calculated first.

\section{For $\mathbf{0<\Lambda<{1/9M^2}}$}
The systemic approach to find the solution is to first find the roots of the cubic equation, followed by setting up the integral with appropriate limits.
This integral, through appropriate transformations, can be brought to the standard form of elliptic integrals and functions.
\subsubsection*{Roots}
Since the cosmological constant is positive, the integrand can be written as,
\begin{equation}
  \label{integrand_1}
  \sqrt{r^2f(r)}=h\sqrt{r(r_3-r)(r-r_2)(r-r_1)}\quad\text{where}\quad h^2=\frac{\Lambda}{3},\:r_1<0<r_2<r<r_3.
\end{equation}
The roots of the cubic equation $\bar{f}(r)=0$, after simplification, reduce to (see Appendix \ref{cubic})
\begin{equation}
  \label{roots_1}
  r_1=-\frac{2}{\sqrt{3}h} \cos{\sigma},\quad r_2=\frac{2}{\sqrt{3}h} \cos\left(\sigma+\frac{\pi}{3}\right),\quad r_3=\frac{2}{\sqrt{3}h} \cos\left(\sigma-\frac{\pi}{3}\right),
\end{equation}
where\footnote{For reference, $\cos\left(3\sigma\right)=\cos^3\sigma-3\sin^2\sigma\,\cos\sigma$ and $\sin\left(3\sigma\right)=3\cos^2\sigma\,\sin\sigma-\sin^3\sigma.$}
\begin{equation}
  \label{roots_more_1}
  \cos\left(3\sigma\right)=\sqrt{27}{Mh}\quad \text{and} \quad \sin\left(3\sigma\right)=\sqrt{1-27M^2h^2}.
\end{equation}

\subsubsection*{Integral}
Finally, from (\ref{transformation_rule}), $r=r(\rho)$ can be found by evaluating,
\begin{equation}
  \label{integration_1}
  \int^{r}_{r_2} \frac{d{r'}}{\sqrt{{r'}^{\,2} f({r'})}}=\int^{r}_{r_2} \frac{d{r'}}{h\sqrt{{r'}(r_3-{r'})({r'}-r_2)({r'}-r_1)}}=\int^{\rho}_{\rho_2} \frac{d{\rho'}}{{\rho'}}.
\end{equation}
Under the transformation (see for example, equation $(256.00)$ of the Handbook \cite{Byrd1971})\footnote{One can apply the transformation (\ref{transformation_1}) since $$\frac{r_3(r'-r_2)}{r'(r_3-r_2)}=\frac{r_3}{(r_3-r_2)}\left(1-\frac{r_2}{r'}\right)=\sin^2\phi,$$ is an increasing function with values $0$ and $1$ at the boundaries $r'=r_2$ and $r'=r_3$ respectively. In general, a direct approach to evaluating this type of integral is to take
\begin{equation}
  \label{condition_1}
  {r'}=\frac{a_1+a_2\,\mathrm{sn}^2({\psi'},k)}{a_3+a_4\,\mathrm{sn}^2({\psi'},k)}\quad\text{such that}\quad\frac{d{r'}}{h\sqrt{{r'}(r_3-{r'})({r'}-r_2)({r'}-r_1)}}=a_5 d{\psi'}
\end{equation}
where $a_i$'s are constants and $0\leq{\psi'}\leq\mathbb{K}(k)=\mathbb{F}(\pi/2,k)$ \cite{Byrd1971}.}
\begin{equation}
  \label{transformation_1} 
  r'=\frac{r_2\,r_3}{r_3-(r_3-r_2)\sin^2\phi},
\end{equation}
the integral (\ref{integration_1}) reduces to the elliptic integral of the first kind (see equation (\ref{first_kind})):
\begin{equation}
    \label{first_kind_1}
    \ln\left(\frac{\rho}{\rho_2}\right)=\frac{2}{h\sqrt{r_3(r_2-r_1)}}\int\displaylimits_{0}^{\Phi}\frac{d\phi}{\sqrt{1-k^2\sin^2\phi}}=\frac{2}{h\sqrt{r_3(r_2-r_1)}}\mathbb{F}(\Phi,k),
\end{equation}
where
\begin{equation}
  \label{modulus_1}
  k^2=\frac{-r_1(r_3-r_2)}{r_3(r_2-r_1)}\leq 1,\quad\Phi(r)=\sin^{-1}\sqrt{\frac{r_3(r-r_2)}{r(r_3-r_2)}}.
\end{equation}
In terms of Jacobi's inverse elliptic function (see equation (\ref{inverse_sn})),
\begin{equation}
  \label{inverse_sn_1}
  \frac{h\sqrt{r_3(r_2-r_1)}}{2}\;\ln\left(\frac{\rho}{\rho_2}\right)=\mathrm{sn}^{-1}\left(\sin\Phi,k\right).
\end{equation}

\subsubsection*{Solution}
From equation (\ref{inverse_sn_1}), $r=r(\rho)$ can be found in terms of the Jacobian elliptic function:
\begin{equation}
  \label{solution_r_1}
  r=\frac{r_2 r_3}{r_3-(r_3-r_2)\,\mathrm{sn}^2(\psi,k)}\quad\text{where}\quad\psi(\rho)=\frac{h}{2}\sqrt{r_3(r_2-r_1)}\;\mathrm{ln}\left(\frac{\rho}{\rho_2}\right),\;k^2=\frac{-r_1(r_3-r_2)}{r_3(r_2-r_1)}.
\end{equation}
Consequently, the conformal factor is,
\begin{equation}
  \label{solution_F_1}
  F(\rho)=r_3h^2(r_2-r_1)(r_3-r_2)^2\cdot\frac{\mathrm{sn}^2(\psi,k)\,\mathrm{dn}^2(\psi,k)\,\mathrm{cn}^2(\psi,k)}{[r_3-(r_3-r_2)\,\mathrm{sn}^2(\psi,k)]^2}.
\end{equation}
And the index of refraction is,
\begin{equation}
  \label{solution_n_1}
  n(\rho)=\frac{r_2}{h(r_3-r_2)}\sqrt{\frac{r_3}{r_2-r_1}}\cdot\frac{1}{\rho\;\mathrm{sn}(\psi,k)\,\mathrm{dn}(\psi,k)\,\mathrm{cn}(\psi,k)}.
\end{equation}
From equations (\ref{first_kind_1}) and (\ref{modulus_1}), the complete elliptic integral reduces to,
\begin{equation}
  \mathbb{K}(k)=\mathbb{F}\left(\Phi=\frac{\pi}{2},k\right)=\frac{h}{2}\sqrt{r_3(r_2-r_1)}\ln\left(\frac{\rho_3}{\rho_2}\right).
\end{equation}
Thus, the ratio between location of horizons is given by,
\begin{equation}
  \label{ratio_horizon}
  \frac{\rho_3}{\rho_2}=\exp\left[{\frac{2}{{h}\sqrt{r_3(r_2-r_1)}}\mathbb{K}(k)}\right].
\end{equation}

\subsection{Schwarzschild limit}
From equation (\ref{roots_more_1}),
\begin{equation}
  \label{more_schw_1}
  3\sigma=\tan^{-1}\left(\sqrt{\frac{1}{27M^2h^2}-1}\right),\quad \lim_{h\to 0}\;\frac{d\sigma}{dh}=-\sqrt{3}M.
\end{equation}
Furthermore, as $\Lambda\to 0$, $h\to 0$ and from equations (\ref{roots_more_1}), (\ref{roots_1}) and (\ref{solution_r_1}),
\begin{equation}
  \label{to_schw_1}
  \sigma\to \pi/6,\;\,hr_1\to -1,\;\,r_2\to 2M,\;\,hr_3\to 1,\;\,\psi\to\mathrm{ln}\left|\sqrt{\frac{\rho}{\rho_2}}\right|,\;\,k\to 1.
\end{equation}
Here, for $r_2$, both the numerator and denominator go to zero as $h\to 0$, and the limit is evaluated by applying L'Hospital's rule and (\ref{more_schw_1}).
For the unit modulus (i.e., $k=1$), Jacobian elliptic functions turn to hyperbolic functions.
\begin{equation}
  \label{elliptic_to_hyperbolic}
  \mathrm{sn}(\psi,1)=\tanh\psi,\quad\mathrm{cn}(\psi,1)=\mathrm{dn}(\psi,1)=\mathrm{sech}\psi.
\end{equation}
Substituting all these conditions into equations (\ref{solution_r_1}), (\ref{solution_F_1}) and (\ref{solution_n_1}) yield,
  \begin{align}
  r_s(\rho)&=2M\cosh^2\left[\ln\left(\sqrt{\frac{\rho}{\rho_H}}\right)\right]=\frac{M}{2}\frac{\rho}{\rho_H}\left(1+\frac{\rho_H}{\rho}\right)^2, \\
  F_s(\rho)&=\tanh^2\left[\ln\left(\sqrt{\frac{\rho}{\rho_H}}\right)\right]=\left(\frac{1-\rho_H/\rho}{1+\rho_H/\rho}\right)^2, \\
  n_s(\rho)&=\frac{2M}{\rho}\cdot\frac{\cosh^3\left[\ln\left(\sqrt{{\frac{\rho}{\rho_H}}}\right)\right]}{\sinh\left[\ln\left(\sqrt{{\frac{\rho}{\rho_H}}}\right)\right]}=\frac{M}{2\rho_H}\left(1+\frac{\rho_H}{\rho}\right)^3\left(1-\frac{\rho_H}{\rho}\right)^{-1}, \label{schw_refractive}
  \end{align}
where the subscript $s$ indicates the Schwarzschild and $H$ horizon.
Since this spacetime is asymptotically flat, $n_s(\rho)\to 1$ as $\rho\to \infty$, yielding $\rho_H=M/2$.
Thus,
\begin{equation}
  \label{schw_refractive_more}
  n_s(\rho)=\left(1+\frac{\rho_H}{\rho}\right)^3\left(1-\frac{\rho_H}{\rho}\right)^{-1}.
\end{equation}
It is easy to see that the refractive indices (\ref{solution_n_1}) and (\ref{schw_refractive}) are not proportional; therefore, in isotropic static coordinates, the Fermat metrics of Kottler and Schwarzschild are not projectively equivalent.

\subsection{Refractive Index}
The Kottler spacetime, unlike the Schwarzschild, is \textit{not} asymptotically flat. 
Additionally, this spacetime in isotropic static form cannot be matched at the boundaries $\rho=\rho_2$ and $\rho=\rho_3$ due to the presence of horizons (from equations (\ref{solution_F_1}) and (\ref{solution_n_1}), $g_{\rho\rho}=F(\rho)n^2(\rho)$ is finite at the horizons, but $g^{tt}=-1/F(\rho)$ is not).
Therefore, one of the horizons, the integration constant, cannot be determined in terms of $M$ and $\Lambda$.

\begin{figure}[h!]
  \caption{
The refractive index (\ref{solution_n_1}) can be written as a function of $\mathrm{ln}\left(\rho/\rho_2\right)$ with the parameter $y=M^2h^2=M^2\Lambda/3$ by utilizing the relationship (\ref{horizon_1}). Here, $0<y<1/27$. Although for the relationship (\ref{horizon_1}) the refractive index depends only on $z$ and $y$, the horizon $\rho_2$ depends upon both $M$ and $h$.
Here, the vertical dashed line represents second horizon and horizontal, refractive index equal to one.
}
  \begin{center}
      \label{Two_horizons}
      \includegraphics[scale=0.6]{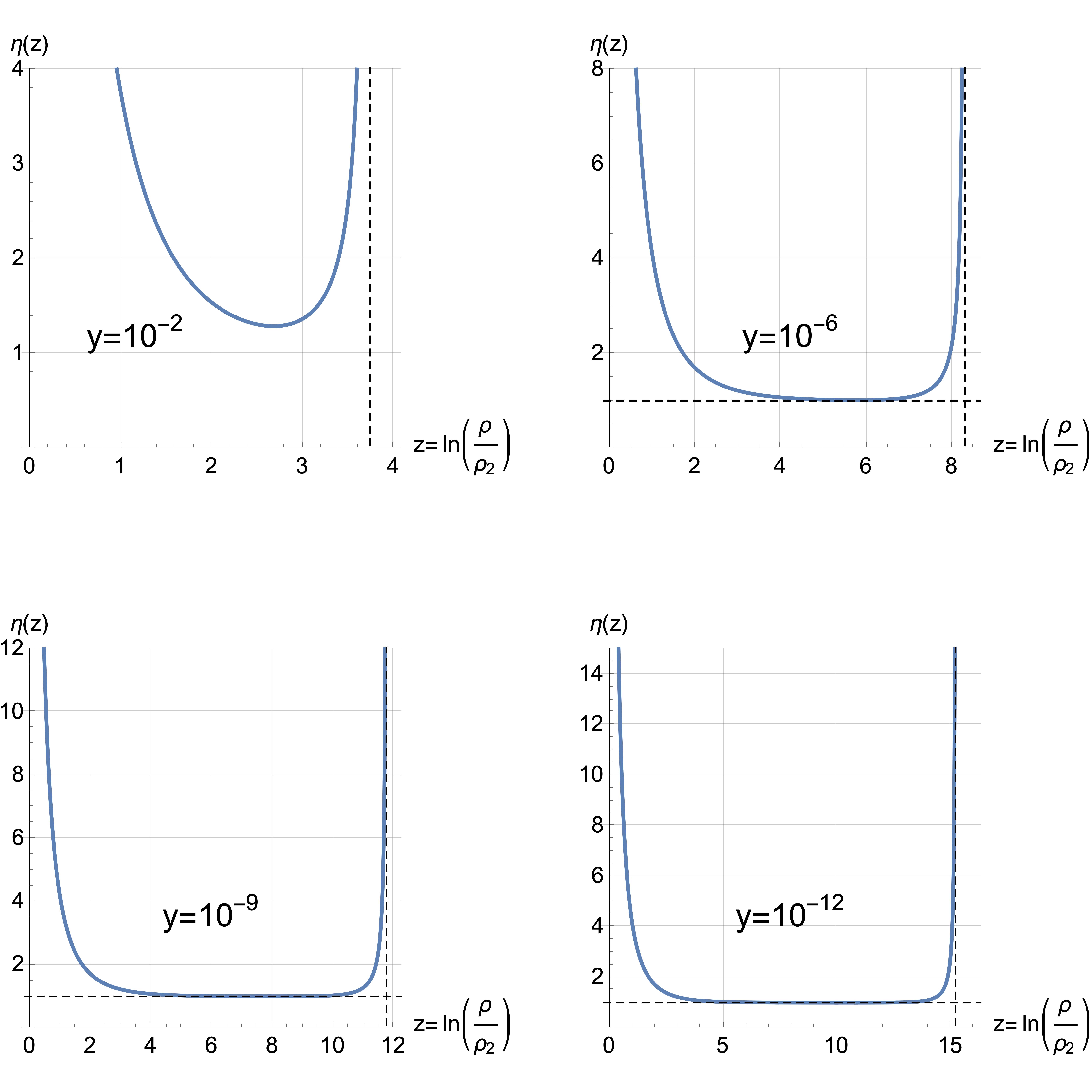}
  \end{center}
\end{figure}

In order to plot the refractive index, we consider it as a function of $z=\ln(\rho/\rho_2)$.
Therefore, from equation (\ref{solution_n_1})
\begin{equation}
  \eta(z)=\frac{r_2}{h(r_3-r_2)}\sqrt{\frac{r_3}{r_2-r_1}}\cdot\frac{1}{\rho_2}\cdot\frac{1}{e^{z}\;\mathrm{sn}(\Psi,k)\,\mathrm{dn}(\Psi,k)\,\mathrm{cn}(\Psi,k)},
\end{equation}
where $2\Psi(z)=z\,h\sqrt{r_3(r_2-r_1)}$.
Without loss of generality, if we further consider 
\begin{equation}
  \label{horizon_1}
  \rho_2=\frac{r_2}{4h(r_3-r_2)}\sqrt{\frac{r_3}{r_2-r_1}},
\end{equation}
then from (\ref{to_schw_1}), $\rho_2\to M/2$ as $\Lambda\to 0$ and
\begin{equation}
  \label{plot_ref_index_1} 	
  \eta(z)=\frac{4}{e^{z}\;\mathrm{sn}(\Psi,k)\,\mathrm{dn}(\Psi,k)\,\mathrm{cn}(\Psi,k)}.
\end{equation}
It is easy to see that $\sigma$, from equation (\ref{roots_more_1}), depends upon a single parameter $y=M^2h^2=M^2\Lambda/3$ where $0<y<1/27$.
Therefore, $hr_i$ (where $i=1,2$ and $3$), modulus $k$ and argument $\Psi$ all depend upon single parameter $y$.
The refractive indices for different values of $y$ are plotted in figure \ref{Two_horizons}.

\section{For $\mathbf{\Lambda<0}$}
This section follows a procedure similar to that in the previous section, except for the compactness of notations, the modulus $k$ in Jacobian elliptic functions is dropped as is usually done.
\subsubsection*{Roots}
Since the cosmological constant is negative, the integrand can be written as,
\begin{equation}
  \label{integrand_2}
  \sqrt{r^2f(r)}=H\sqrt{r(r-r_1)(r-r_2)(r-r_3)}\quad\text{where}\quad H^2=-\frac{\Lambda}{3},\:0<r_1<r<\infty.
\end{equation}
The roots of the cubic equation $\bar{f}(r)=0$, after simplification, reduces to (see Appendix \ref{cubic})
\begin{equation}
  \label{roots_2}
  r_1=S_1+S_2,\quad \mathrm{\mathbb{R}e}(r_2)=\mathrm{\mathbb{R}e}(r_3)=-\frac{r_1}{2},\quad \mathrm{\mathbb{I}m}(r_2)=-\mathrm{\mathbb{I}m}(r_3)=\frac{\sqrt{3}}{2}(S_1-S_2),
\end{equation}
where $\mathrm{\mathbb{R}e}$ and $\mathrm{\mathbb{I}m}$ represent the real and imaginary parts of a complex number respectively and
\begin{equation}
  \label{roots_more_2}
  S_1=\frac{1}{H}\left(MH+\sqrt{\frac{1}{27}+M^2H^2}\right)^{{1/3}},\quad S_2=\frac{1}{H}\left(MH-\sqrt{\frac{1}{27}+M^2H^2}\right)^{{1/3}}.
\end{equation}

\subsubsection*{Integral}
Thus, $r=r(\rho)$ can be found by evaluating,
\begin{equation}
  \label{integration_2}
  \int^{r}_{r_1} \frac{d{r'}}{\sqrt{{r'}^2 f({r'})}}=\int^{r}_{r_1} \frac{d{r'}}{H\sqrt{{r'}({r'}-r_1)({r'}-r_2)({r'}-r_3)}}=\int^{\rho}_{\rho_1} \frac{d{\rho'}}{{\rho'}}.
\end{equation}
This integration can be solved in a similar fashion as in the previous section, except in equation (\ref{condition_1}) $\mathrm{sn}({\psi'},k)$ turns to $\mathrm{cn}({\psi'},k)$ since $r_2$ and $r_3$ form complex conjugate pair (see for example, equation $(260.00)$ of the Handbook \cite{Stegun}).
However, the resultant modulus $k$ is greater than one.
One can apply reciprocal modulus transformation, such that the modulus converts to $k_1=1/k$.
However, the Jacobian amplitude becomes imaginary (or the cosine of the amplitude becomes greater than one).
One way to circumvent these problems is to apply the transformation $x=1/r$ which leads to a polynomial of degree three under square root (this will lead to compactifiation of space however):
\begin{equation}
  \mathrm{ln}\left(\frac{\rho}{\rho_1}\right)=\frac{1}{\sqrt{2M}} \int^{x_1}_{x}\frac{d{x'}}{\sqrt{(x_1-{x'})(x_2-{x'})(x_3-{x'})}}\quad \text{where}\quad x_i={1 \over r_i},\, 0<x<x_1
\end{equation}
This integral can be evaluated, in general, by substitution (see for example, equation $(243.00)$ of the Handbook \cite{Byrd1971} for this case),
\begin{equation}
  \label{condition_2}
  {x'}=\frac{a_1+a_2\,\mathrm{cn}({\psi'},k)}{a_3+a_4\,\mathrm{cn}({\psi'},k)}\quad\text{such that}\quad\frac{d{x'}}{\sqrt{(x_1-{x'})(x_2-{x'})(x_3-{x'})}}=a_5\,d{\psi'}.
\end{equation}
where $a_i$'s are constants, ${\psi'}={\psi'}({\rho'}$) and $0\leq{\psi'}\leq 2\mathbb{K}(k)$.

\subsubsection*{Solution}
All constants ($a_i$'s) from equation (\ref{root_properties}) can be evaluated in terms of $r_1$, $M$ and $H$. After the simplification,
\begin{equation}
  \label{solution_r_2}
  r=\frac{r_1\sqrt{M}\,(1+\mathrm{cn}\psi)}{\sqrt{M}-\sqrt{3M-r_1}+(\sqrt{M}+\sqrt{3M-r_1})\,\mathrm{cn}\psi},
\end{equation}
where
\begin{equation}
  \label{solution_mod_2}
  \psi(\rho)=\left[\frac{4M\,(3M-r_1)}{r_1^2}\right]^{1/4}\mathrm{ln}\left(\frac{\rho}{\rho_1}\right),\quad k^2=\frac{1}{2}+\frac{6M-r_1}{\sqrt{64M(3M-r_1)}}\leq 1.
\end{equation}
Now, as $r\to\infty$, from the equation (\ref{solution_r_2})
\begin{equation}
  \label{range_1}
  \mathrm{cn}\psi_\infty=\frac{\sqrt{3M-r_1}-\sqrt{M}}{\sqrt{3M-r_1}+\sqrt{M}}=\phi_\infty\implies \psi_\infty=\mathrm{cn}^{-1}\left(\phi_\infty,k\right)=\mathbb{F}\left[\cos^{-1}(\phi_\infty),k\right],
\end{equation}
yielding the range,
\begin{equation}
  \label{range_2}
  \rho\in (\rho_1,\rho_\infty)\quad\text{where}\quad\frac{\rho_\infty}{\rho_1}=\exp\left[\psi_\infty\,\left(\frac{r_1^2}{4M\,(3M-r_1)}\right)^{1/4} \right].
\end{equation}
Thus, the transformation $x=1/r$ has compactified the space.
Moreover, the solution is finite (i.e., the Jacobian functions involved in the solution are non-zero), since the complete integral $\mathbb{K}(k)$ occurs for $r<0$ and $r<r_1$ i.e., outside the domain.
Finally, the conformal factor is given by,
\begin{equation}
  \label{solution_F_2}
  F(\rho)=8\,\sqrt{{M r_1}\left(\frac{3M}{r_1}-1\right)^3}\cdot\frac{\mathrm{dn}^2\psi\,(1-\mathrm{cn}\psi)}{{(1+\mathrm{cn}\psi)}\left[\sqrt{M}-\sqrt{3M-r_1}+(\sqrt{M}+\sqrt{3M-r_1})\,\mathrm{cn}\psi\right]^2}.
\end{equation}
As expected, the conformal factor $F\left(\rho_\infty\right)\to\infty$, the requirement for conformal compactification (see for example, Chapter $2.4.1$ of Ref.\cite{Townsend:1997ku} and references within).
The index of refraction,
\begin{equation}
  \label{solution_n_2}
  n(\rho)=\left(\frac{Mr_1^6}{64\,(3M-r_1)^3}\right)^{1/4} \cdot\frac{\left(1+\mathrm{cn}\psi\right)^2}{\rho\,\mathrm{sn}\psi\,\mathrm{dn}\psi}\:,
\end{equation}
shows that the projective equivalence no longer exist under transformation to isotropic coordinates.

\subsection{Schwarzschild limit}
As $\Lambda\to 0$, $H\to 0$ and from equations (\ref{roots_2}), (\ref{roots_more_2}) and (\ref{solution_mod_2})
\begin{equation}
  r_1\to 2M,\quad k\to1,\quad \psi\to\mathrm{ln}\left(\frac{\rho}{\rho_1}\right),
\end{equation}
and since Jacobian elliptic functions turn to hyperbolic functions (see equation (\ref{elliptic_to_hyperbolic})), the Schwarzschild solution is recovered.
Moreover, asymptotic flatness yields $\rho_1\to\rho_H=M/2$ and since $\mathbb{K}(k=1)=\mathbb{F}(\pi/2,1)\to \infty$, from equation (\ref{range_2}), $\rho_\infty\to \infty$.

\subsection{Refractive Index}
\begin{figure}[h!]
  \caption{
  The refractive indices are plotted as a function of $\rho/\rho_1$ using the relationship (\ref{horizon_2})
  The vertical dashed line represents the location where $\rho=\rho_\infty$ and horizontal, the refractive index at $\rho_\infty$. The asymptotic value of refractive index can be found, since the spacetime has only one horizon and the space has been compactified. Another horizontal line, dot dashed, indicates unit index of refraction.
  The parameter $y=M^2H^2=-M^2\Lambda/3$, when less than one, gives rise to index of refraction less than one.
  However, if the relationship (\ref{horizon_2}) is not applied, then the index of refraction may depend upon both $M$ and $H$ (rather than only on $y$) such that it is greater than one.}
  \begin{center}
      \label{One_horizon}	
      \includegraphics[scale=0.52]{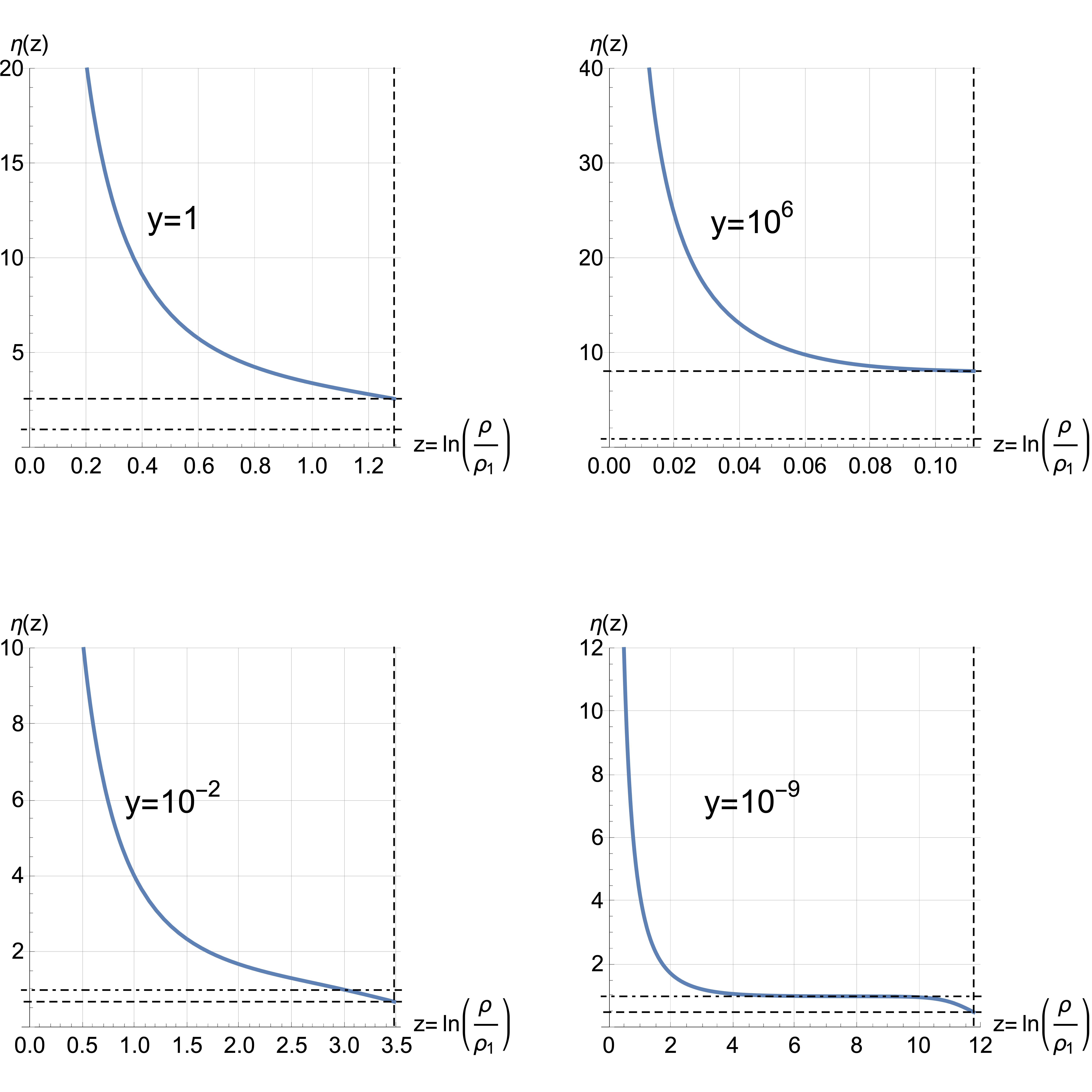}
  \end{center}
\end{figure}
Once again, to plot the refractive index, we consider it as a function of $z=\ln(\rho/\rho_1)$. Therefore, from equation (\ref{solution_n_2})
\begin{equation}
     \eta(z)=\left(\frac{Mr_1^6}{64\,(3M-r_1)^3}\right)^{1/4}\cdot \frac{1}{\rho_1} \cdot\frac{\left(1+\mathrm{cn}\Psi\right)^2}{e^{z}\,\mathrm{sn}\Psi\,\mathrm{dn}\Psi}\:,
\end{equation}
where
\begin{equation}
      \Psi(z)=z\left[\frac{4M\,(3M-r_1)}{r_1^2}\right]^{1/4}.
\end{equation}
Without loss of generality, if we consider
\begin{equation}
  \label{horizon_2}
  \rho_1=\frac{1}{2}\left(\frac{Mr_1^6}{64\,(3M-r_1)^3}\right)^{1/4}.
\end{equation}
then $\rho_1\to M/2$ as $\Lambda\to 0$, and
\begin{equation}
     \eta(z)=\frac{2\,\left(1+\mathrm{cn}\Psi\right)^2}{e^{z}\,\mathrm{sn}\Psi\,\mathrm{dn}\Psi}\,.
\end{equation}
From equations (\ref{roots_more_2}) and (\ref{roots_2}), it is easy to see that $Hr_1$ and in turn amplitude $\Psi(z)$ and modulus $k$ depend upon a single parameter $y=M^2H^2=-M^2\Lambda/3$.
The refractive indices for different values of $y$ are plotted in figure \ref{One_horizon}.

\section{Conclusion}

For the Kottler spacetime, the static region exists between the two horizons when $0<\Lambda<1/9M^2$, and beyond a single horizon when $\Lambda<0$.
For these static regions, we have calculated the metric in isotropic coordinates.
For $\Lambda<0$, the refractive index can be found asymptotically, as the space is conformally compactified and the static region extends to spatial infinity.

In static coordinates, the spatially projected (affinely parameterized) null geodesics of a Lorentzian spacetime are the (non-affinely parameterized) geodesics of the associated Riemannian Fermat metric.
As mentioned in the introduction, the parameterization changes due to the conformal transformations.
The property that the unparameterized geodesics of two Riemannian metrics coincide is called the projective equivalence. And a necessary condition for this to occur is that both metrics share the identical projective curvature tensor.
In isotropic static coordinates, we have shown that the necessary and sufficient condition for the projective equivalence between two Fermat metrics is that the ratio of refractive indices is constant.
In other words, the projective equivalence in isotropic coordinates is analogous to the invariance of Snell's law in ordinary geometric optics.

The Nariai spacetime is a spherically symmetric static solution of the Einstein field equations with cosmological constant and empty matter distribution, i.e., $R_{\mu\nu}=\Lambda g_{\mu\nu}$.
The canonical form for this spacetime does not exist; however, the line element in isotropic coordinates is given by \cite{Nariai_1999},
\begin{equation}
  \label{Nariai}
	ds^2=\frac{1}{\Lambda}\left[-\left(A_1\cos(\ln\rho)+A_2\sin(\ln\rho)\right)^2 dt^2+\frac{1}{\rho^2}\left(d\rho^2+\rho^2d\Omega^2\right)\right].
\end{equation}
As can be seen, an attempt to transform to the canonical form due to the presence of $\rho^{-2}$ results in $r$ being constant. 
Since the null geodesics retain affine parameterization if the metric is multiplied by a constant conformal factor \cite{Wald:1984rg}, the affinely parameterized null geodesics of the Nariai metric (\ref{Nariai}) are independent of $\Lambda$.
It is interesting to note that the refractive indices of the Nariai, Schwarzschild and Kottler spacetimes follow trigonometric, hyperbolic (equation (\ref{schw_refractive})) and elliptic (Jacobian) functions respectively with the natural log of the radial coordinate as the argument.
It is known that the Fermat metrics of the Kottler and Schwarzschild spacetimes in canonical coordinates are projectively equivalent.
We have shown that this projective equivalence is lost when transformed to isotropic static coordinates which leads to spatially projected null geodesics that are different.
In other words, for the Kottler spacetime in canonical coordinates and Nariai in isotropic, both of which satisfy the field equations $R_{\mu\nu}=\Lambda g_{\mu\nu}$, the spatial trajectories of affinely parameterized null geodesics are independent of $\Lambda$.

\section*{Acknowledgement}
I thank my supervisor, Dr. Mohammad Akbar, for his excellent course \textit{Theory of Black Holes} and for introducing me to the fascinating area of projective equivalence and gravitational lensing with Noah Bright from Pittsburgh during his REU at UTD in summer 2020. I also thank him for his guidance and sharing his insights from his work with Behshid Kasmaie on the subject. I am grateful to Dr. Gary Gibbons for his encouraging and insightful comments. I thank Noah for stimulating exchanges and for being my first mentee. Special thanks to Charlie Brewer for a careful reading of the manuscript. I also thank the anonymous referees for their useful suggestions. I have been partially supported by Julia Williams Van Ness Merit Scholarship and Margie Renfrow Student Funds.

\appendix
\section{ Elliptic integrals and functions}
\label{elliptic}
This appendix lists some of the fundamental formulae related to elliptic integrals and functions (see \cite{Byrd1971} for more details).
The integral
\begin{equation}
I=\int R\left[x, \sqrt{p(x)}\right] dx,
\end{equation}
is known as elliptic integral if $p(x)$ is a polynomial of degree four or three with no multiple roots.
Here, $R$ is a rational function of $x$ and $\sqrt{p(x)}$.
It was shown by Legendre that any elliptic integral can be expressed linearly in terms of elementary functions and three fundamental integrals known as the canonical elliptic integrals of first, second and third kind.
The elliptic integral of the first kind
\begin{equation}
  \label{first_kind}
  \mathbb{F}(\psi,k)=\int^{\psi}_{0}\frac{d{{\psi'}}}{\sqrt{1-k^2\sin^2{\psi'}}}=\int^{y}_{0}{dy' \over \sqrt{(1-y'^{\,2})(1-k^2y'^{\,2})}}
\end{equation}
is of prime importance to this paper.
Here $k$ is called the modulus ($0<k<1$) and $\psi$ (or $y$) is called the argument ($0<\psi\leq\pi/2$ or $0<y\leq 1$).
Moreover, $y'=\sin\psi'$.
When $\psi=\pi/2$ (or $y=1$), it is known as complete elliptic integral,
\begin{equation}
  \mathbb{K}(k)=\int^{\frac{\pi}{2}}_{0}\frac{d{{\psi'}}}{\sqrt{1-k^2\sin^2{\psi'}}}\,.
\end{equation}
The elliptic functions of Jacobi are obtained through inversion of elliptic integrals of the first kind.
\begin{eqnarray}
  \mathrm{am}^{-1} (\psi,k)&=&\int^{\psi}_{0}\frac{d\psi'}{\sqrt{1-k^2\sin^2\psi'}}=\mathbb{F}(\psi,k)=u, \\
  \mathrm{sn}^{-1} (y,k)&=&\int^{y}_{0}\frac{dy'}{\sqrt{(1-y'^{\,2})(1-k^2y'^{\,2})}}=\mathbb{F}(\sin^{-1}y,k)=u. \label{inverse_sn}
\end{eqnarray}
Thus, $y=\sin\psi=\mathrm{sn}(u,k)$ and $\psi=\mathrm{am}(u,k)$.
Here, $\mathrm{sn}(u,k)$ may be read \textit{sine amplitude u} and $\mathrm{am}(u,k)$, \textit{amplitude u}.
Moreover,
\begin{eqnarray}
  \mathrm{cn}^{-1} (y,k)&=&\int^{1}_{y}\frac{dy'}{\sqrt{(1-y'^{\,2})(K^{\,2}+k^2y'^{\,2})}}=\mathbb{F}(\sin^{-1}\sqrt{1-y^{\,2}},k), \\
  \mathrm{dn}^{-1} (y,k)&=&\int^{1}_{y}\frac{dy'}{\sqrt{(1-y'^{\,2})(y'^{\,2}-K^{\,2})}}=\mathbb{F}\left(\sin^{-1}\sqrt{(1-y^2)/k^2},k\right),\; K\leq y< 1
\end{eqnarray}
where $K=\sqrt{1-k^2}$ is called the complementary modulus.
Moreover, the range is given by
\begin{equation}
  -1\leq \mathrm{sn}\psi\leq 1,\quad -1\leq \mathrm{cn}\psi\leq 1,\quad K\leq \mathrm{dn}\psi\leq 1,
\end{equation}
and special values,
\begin{equation}
     \mathrm{am}\,\mathbb{K}=\pi/2,\quad \mathrm{sn}\,\mathbb{K}=1,\quad \mathrm{cn}\,\mathbb{K}=0,\quad \mathrm{dn}\,\mathbb{K}=K,\quad \mathrm{am}\,0=0,\quad \mathrm{sn}\,0=0,\quad \mathrm{cn}\,0=1,\quad \mathrm{dn}\,0=1.
\end{equation}
The Jacobian elliptic functions turn to hyperbolic when the modulus $k=1$
\begin{equation}
  \mathrm{sn}(\psi,1)=\tanh\psi,\quad\mathrm{cn}(\psi,1)=\mathrm{sech}\psi,\quad\mathrm{dn}(\psi,1)=\mathrm{sech}\psi,
\end{equation}
and to trigonometric when it vanishes,
\begin{equation}
  \mathrm{sn}(\psi,0)=\sin\psi,\quad\mathrm{cn}(\psi,0)=\cos\psi,\quad\mathrm{dn}(\psi,0)=1.
\end{equation}
Additionally, differentiation with respect to the argument is given by,
\begin{equation}
  \frac{\partial}{\partial \psi} (\mathrm{am}\psi)=\mathrm{dn}\psi,\quad \frac{\partial}{\partial \psi} (\mathrm{sn}\psi)=\mathrm{cn}\psi\,\mathrm{dn}\psi,\quad \frac{\partial}{\partial \psi} (\mathrm{cn}\psi)=-\mathrm{sn}\psi\,\mathrm{dn}\psi,\quad \frac{\partial}{\partial \psi} (\mathrm{dn}\psi)=-k^2\mathrm{sn}\psi\,\mathrm{cn}\psi.
\end{equation}

\section{Roots of Cubic Equation}
\label{cubic}
The roots $x_1$, $x_2$ and $x_3$ of the cubic equation
\begin{equation}
  x^3+a_2x^2+a_1x+a_0=0,
\end{equation}
are given by \cite{Stegun}
\begin{align}
    x_1&=S_1+S_2-\frac{a_2}{3}, \\
    x_2&=-\frac{1}{2}\left(S_1+S_2\right)-\frac{a_2}{3}+\frac{i\,\sqrt{3}}{2}\left(S_1-S_2\right), \\
    x_3&=-\frac{1}{2}\left(S_1+S_2\right)-\frac{a_2}{3}-\frac{i\,\sqrt{3}}{2}\left(S_1-S_2\right),
\end{align}
where
\begin{equation}
  S_1=\left[p+\left(q^3+p^2\right)^{\frac{1}{2}}\right]^{\frac{1}{3}},\quad S_2=\left[p-\left(q^3+p^2\right)^{\frac{1}{2}}\right]^{\frac{1}{3}},
\end{equation}
and
\begin{equation}
  q=\frac{a_1}{3}-\left(\frac{a_2}{3}\right)^2,\quad p=\frac{1}{6}\left(a_1a_2-3a_0\right)-\left(\frac{a_2}{3}\right)^3.
\end{equation}
Moreover, if
\begin{equation}
  q^3+p^2 \quad
   \left\{
   \begin{aligned}
     &>0, \text{ then a real root and a complex conjugate pair.} \\
     &=0, \text{ then all roots are real and at least two are equal.} \\
     &<0, \text{ then all roots are real.}
   \end{aligned}
   \right.
\end{equation}

\section{Light Rays in Isotropic Coordinates}
\label{Light Rays in Isotropic Coordinates}
The geodesics of the Fermat metric (\ref{optical_metric}) can be derived from the Lagrangian,
\begin{equation}
  \mathcal{L}=\frac{n^2(\rho)}{2}\left(\dot{\rho}^2+\rho^2\,\dot{\theta}^2+\rho^2\sin^2\theta\,\dot{\phi}^2\right),
\end{equation}
where $\dot{\rho}=d\rho/d\lambda$ and $\lambda$ is affine parameter.
The Euler-Lagrange equation in $\theta$ reduces to,
\begin{equation}
  \ddot{\theta}=\dot{\phi}^2\sin\theta\cos\theta-2\,\dot{\rho}\,\dot{\theta}\left(\frac{\dot{n}}{n}+\frac{1}{\rho}\right),
\end{equation}
where $\dot{n}=dn/d\rho$.
For the initial conditions, $\theta=\pi/2$ and $\dot{\theta}=0$, the equation above yields $\ddot{\theta}=0$.
Thus, $\theta$ remains $\pi/2$ and the motion is confined to the equatorial plane.
Since the metric is rotationally invariant, we can consider motion in equatorial plane without loss of generality.
Therefore, the Euler-Lagrange equations in $\rho$ and $\phi$ reduces to
\begin{equation}
  \label{ELE_rho}
  \ddot{\rho}+ \left(\frac{\dot{n}}{n}\right) \dot{\rho}^2=\rho\,\dot{\phi}^2 + \rho^2 \left(\frac{\dot{n}}{n}\right) \dot{\phi}^2,
\end{equation}
\begin{equation}
  \label{ELE_phi}
  \ddot{\phi}+2\,\dot{\rho}\,\dot{\phi} \left(\frac{\dot{n}}{n}+\frac{1}{\rho}\right)=0.
\end{equation}
Since we are interested in unparameterized trajectory i.e., $\rho=\rho(\phi)$, from the chain rule,
\begin{equation}
  \frac{d\rho}{d\lambda}=\frac{d\rho}{d\phi}\;\frac{d\phi}{d\lambda},
\end{equation}
and,
\begin{equation}
  \frac{d^2\rho}{d\lambda^2}=\frac{d\rho}{d\phi}\;\frac{d^2\phi}{d\lambda^2}+\frac{d^2\rho}{d\phi^2}\left(\frac{d\phi}{d\lambda}\right)^2.
\end{equation}
Substituting these relations into (\ref{ELE_rho}) and (\ref{ELE_phi}), upon simplification, yields
\begin{equation}
  \frac{d^2\rho}{d\phi^2}=\left(\frac{d\rho}{d\phi} \right)^2 \left[\frac{\dot{n}}{n}+\frac{2}{\rho} \right]+\rho+\rho^2 \left(\frac{\dot{n}}{n}\right),
\end{equation}
which is unparameterized spatial trajectory of null geodesics of metric (\ref{isotropic_metric}) in equatorial plane.


\begin{thebibliography}{99}

\bibitem{Akbar_Behshid2021}
  M.~M.~Akbar and B.~Kasmaie,
  ``Gravitational Lensing in Static Spherically Symmetric Spacetimes: Symmetries and Equivalences,"
  \textit{forthcoming}.

\bibitem{Abramowicz88}
M.~A.~Abramowicz, B.~Carter and J.~P.~Lasota,
``Optical Reference Geometry for Stationary and Static Dynamics,"
Gen. Relat. Gravit. \textbf{20}, no. 11, pp. 1173--1183 (1988).

\bibitem{Stegun}
M.~Abramowitz and I.~A.~Stegun (1972)
\textit{Handbook of Mathematical Functions with Formulas, Graphs, and Mathematical Tables}
(United States National Bureau of Standards).

\bibitem{Byrd1971}
P.~F.~Byrd and M.~D.~Friedman (1971)
\textit{Handbook of Elliptic Integrals for Engineers and Scientists}
(Second Edition, Revised, Springer-Verlag Berlin Heidelberg).

\bibitem{Dodelson:2003ft}
S.~Dodelson (2003)
 \textit{Modern Cosmology}
(Academic Press, Amsterdam).

\bibitem{Eisenhart1926}
L.~P.~Eisenhart (1926)
\textit{Riemannian Geometry} (Princeton University Press, Princeton).

\bibitem{Gibbons:2008ru}
G.~W.~Gibbons, C.~M.~Warnick and M.~C.~Werner,
``Light-bending in Schwarzschild-de-Sitter: Projective Geometry of the Optical Metric,''
Class. Quant. Grav. \textbf{25}, 245009 (2008).

\bibitem{Islam:1983rxp}
J.~Islam,
``The Cosmological Constant and Classical Tests of General Relativity,''
Phys. Lett. A \textbf{97}, 239-241 (1983).

\bibitem{Lakshmi2002}
V.~Lakshminarayanan, A.~Ghatak, Ajoy, Thyagarajan, K. (2002)
 \textit{Lagrangian Optics}
 (Springer US).

\bibitem{Landau:1982dva}
L.~D.~Landau and E.~M.~Lifschits (1975)
\textit{The Classical Theory of Fields}
(Fourth Revised English Edition, Elsevier).

\bibitem{McVittie:1933zz}
G.~C.~McVittie,
``The mass-particle in an expanding universe,''
Mon. Not. Roy. Astron. Soc. \textbf{93}, 325-339 (1933).

\bibitem{Nariai_1999}
H.~Nariai,
``On a New Cosmological Solution of Einstein's Field Equations of Gravitation,"
Gen. Relat. Gravit. \textbf{31}, 963--971 (1999).

\bibitem{Perlick:2010zh}
V.~Perlick,
``Gravitational Lensing from a Spacetime Perspective,''
\textit{Living Rev. Relativ.} \textbf{7}, 9 (2004)

\bibitem{Perlick_book}
V.~Perlick (2000)
\textit{Ray optics, Fermat's principle and applications to general relativity}
(Springer, Heidelberg).

\bibitem{EFS}
P.~Schneider, J.~Ehlers and E.~.E.~Falco (1992)
\textit{Gravitational Lenses}
(Springer, Berlin).

\bibitem{SS}
J.~L.~Synge and A.~Schild (1978)
\textit{Tensor Calculus}
(Dover publications, New York).

\bibitem{Townsend:1997ku}
P.~K.~Townsend,
``Black Holes: Lecture Notes,''
arXiv:gr-qc/9707012

\bibitem{Wald:1984rg}
R.~M.~Wald (1984)
\textit{General Relativity}
(University of Chicago Press, Chicago).

\end{thebibliography}
\end{document}